\documentclass[sigconf]{acmart}

\usepackage[utf8]{inputenc}
\usepackage{bm}
\usepackage{balance}
\usepackage{booktabs} 
\usepackage{graphics}
\usepackage{multirow}
\usepackage{array}
\usepackage{flushend}
\usepackage[english]{babel}
\usepackage[T1]{fontenc}
\usepackage{textcomp}
\usepackage{latexsym}
\usepackage{natbib}
\usepackage{graphicx}
\usepackage{multirow}
\usepackage[labelformat=simple]{subcaption}
\usepackage{algorithm}
\usepackage{algorithmic}

\newcommand{\dou}[1]{}
\newcommand{\zhou}[1]{}
\begin{document}
\title{PSSL: Self-supervised Learning for Personalized Search \\ with Contrastive Sampling}
\author{Yujia Zhou$^{2}$, Zhicheng Dou$^{1*}$, Yutao Zhu$^{3}$, and Ji-Rong Wen$^{4,5}$}

\copyrightyear{2021}
\acmYear{2021}
\setcopyright{acmlicensed}\acmConference[CIKM '21]{Proceedings of the 30th ACM International Conference on Information and Knowledge Management}{November 1--5, 2021}{Virtual Event, QLD, Australia}
\acmBooktitle{Proceedings of the 30th ACM International Conference on Information and Knowledge Management (CIKM '21), November 1--5, 2021, Virtual Event, QLD, Australia}
\acmPrice{15.00}
\acmDOI{10.1145/3459637.3482379}
\acmISBN{978-1-4503-8446-9/21/11}

\affiliation{%
  $^1$Gaoling School of Artificial Intelligence, Renmin University of China \\
  $^2$School of Information, Renmin University of China \\
  $^3$Université de Montréal, Montréal, Québec, Canada \\
  $^4$Beijing Key Laboratory of Big Data Management and Analysis Methods \\
  $^5$Key Laboratory of Data Engineering and Knowledge Engineering, MOE
 \country{}
}
\email{{zhouyujia, *dou}@ruc.edu.cn}
\begin{abstract}
Personalized search plays a crucial role in improving user search experience owing to its ability to build user profiles based on historical behaviors. Previous studies have made great progress in extracting personal signals from the query log and learning user representations. However, neural personalized search is extremely dependent on sufficient data to train the user model. Data sparsity is an inevitable challenge for existing methods to learn high-quality user representations. Moreover, the overemphasis on final ranking quality leads to rough data representations and impairs the generalizability of the model. To tackle these issues, we propose a \textbf{P}ersonalized \textbf{S}earch framework with \textbf{S}elf-supervised \textbf{L}earning (PSSL) to enhance data representations. Specifically, we adopt a contrastive sampling method to extract paired self-supervised information from sequences of user behaviors in query logs. Four auxiliary tasks are designed to pre-train the sentence encoder and the sequence encoder used in the ranking model. They are optimized by contrastive loss which aims to close the distance between similar user sequences, queries, and documents. Experimental results on two datasets demonstrate that our proposed model PSSL achieves state-of-the-art performance compared with existing baselines. 
\end{abstract}


\def\authors{Yujia Zhou, Zhicheng Dou, Yutao Zhu, and Ji-Rong Wen}
\settopmatter{printacmref=true}
\maketitle

\section{Introduction}\label{sec:intro}
Search engines have been widely employed across the world as a common tool for retrieving information. For the same query, these platforms usually return the same document list for all users. This strategy is difficult to meet the needs of all users due to the diversity of user interests. To cope with this problem, personalized search has been proposed to re-rank the search results to meet user's individual needs~\cite{teevan2008personalize, dou2007large}. One essential problem in personalized search is how to model the user preferences with respect to his query log. Some early studies~\cite{teevan2011understanding, harvey2013building, Vu2014Improving, vu2015temporal} tried to extract personalized features from user's click-through data to predict user interests. Recently, deep learning based methods~\cite{Lu:2019, ZhouDW20, YaoDXW20, sigir/ZhouDW20, sigir/YaoDW20} have been proposed to build user profiles in semantic space. They applied neural networks to learn effective user representations and brought significant improvements in user satisfaction.

Although existing neural methods have made great progress in improving user search experience, there are two weaknesses that limit their ability to build user profiles. \textbf{First}, these models rely on large amounts of training data to learn user representations more accurately. However, some users only have limited click-through data to train the user representations. Data sparsity is a major challenge for neural personalized search models. \textbf{Second}, existing approaches optimize the model with only the personalized ranking task, which will make the model overemphasize the final ranking quality. The characteristics of query logs, such as the correlations between user behaviors, are not well captured in data representations. Such rough data representations will damage the generalizability of the model. In fact, better data representations can further improve the quality of search results. As can be seen from language models proposed recently~\cite{peters2018deep, devlin2018bert,DBLP:conf/sigir/SuDZQW21,DBLP:conf/aaai/ZhuZNLD21}, the pre-trained data representations benefit the performance of various downstream tasks. This inspires us to rethink the process of model optimization in personalized search. We attempt to learn pre-trained data representations adapting to personalized search, and fine-tune them on the ranking task.

Before the ranking task, how to incorporate the characteristics of personalized search logs in data representations has become a new challenge. Recently, self-supervised learning has achieved great success in various information retrieval tasks, such as sequential recommendation~\cite{cikm/ZhouWZZWZWW20, CL4SR} and ad-hoc document ranking~\cite{iclr/ChangYCYK20, wsdm/MaGZFJC21}. This is a new paradigm for unsupervised representation learning, which aims to learn intrinsic data correlation from the raw data without any supervision signal. The basic process of self-supervised learning is to first construct training samples, and then devise auxiliary tasks to pre-train the model. Its advantages perfectly fit our needs in dealing with the aforementioned two problems: data sparsity and rough data representations. Due to its powerful ability of extracting self-supervised signals from raw data, we intend to apply it to capture behavioral characteristics in query logs, thereby enhancing data representations for personalized search. 

To support personalized search, we need two different categories of representation learning tasks: sentence encoding and sequence encoding. The former aims to learn the representations of queries and documents, while the latter focuses on behavior sequence modeling to obtain the user representations based on their query logs. In fact, the users' query logs contain a lot of paired self-supervised information. For instance, if a user issues two similar queries to find the same document, these two queries reveal the same intention. Moreover, two users who submit the same query to retrieve the same document should show some similarities. To model such paired search patterns in the query logs, we adopt a contrastive sampling method to generate self-supervised signals for pre-training tasks. Under this strategy, we can extract contrastive pairs that show similar meanings from query logs. They help pre-train the parameters of encoders with contrastive learning objectives \cite{icml/ChenK0H20, cvpr/He0WXG20}.

Specifically, we propose a \textbf{P}ersonalized \textbf{S}earch framework with \textbf{S}elf-supervised \textbf{L}earning (PSSL), which is a neural model with two-stage training. \textbf{At the first stage}, we use self-contrastive sampling and user-contrastive sampling to generate self-supervised signals.\dou{four tasks, two types, somewhat ambiguous}\zhou{change their order} The former constructs the contrastive pairs from the query log of a single user, while the latter extracts pairs from different users. Based on the sampling from these two angles, four types of pairs, namely query pairs, document pairs, sequence augmentation pairs, and user pairs, are constructed for self-supervised learning.\dou{are the pairs the same thing with the four tasks mentioned previously?}\zhou{yes} They correspond to four self-supervised tasks to pre-train the sentence encoder and the sequence encoder. As such, pre-trained encoders adapt data representations to personalized search scenarios. \textbf{At the second stage}, two encoders are applied to enhance the personalization, which will be fine-tuned with respect to the ranking quality. To verify the validity of our model, we conduct experiments on search logs from two real-world search engines. Experimental results demonstrate that our model PSSL achieves state-of-the-art performance compared with existing search models.

Our main contributions can be summarized as follows. (1) We propose a two-stage training framework (i.e., pre-training and ranking) for personalized search to strengthen data representations. To the best of our knowledge, this is the first time that the use of pre-training tasks for personalized search is investigated. (2) We use self-supervised learning to capture correlations between user behaviors at the pre-training stage, so as to learn better representations of user behaviors. (3) We adopt a contrastive sampling method to generate training pairs for self-supervised learning. Based on a contrastive learning framework, four self-supervised tasks are devised to pre-train the encoders used in the ranking task.\zhou{re-write the contributions}





\section{Related Work}\label{sec:related_work}
\subsection{Personalized Search}
Personalized search has been a research hotspot because it can improve the ranking quality of search engines effectively~\cite{cai2014personalized}. Traditionally, click-based features are widely studied due to its easy availability and reliability. \citet{dou2007large} proposed P-click and G-click models to count the number of historical clicks on the same query from individual and group behaviors respectively, and fused the results of personalized ranking and original ranking to get the final results. \citet{teevan2011understanding} also collected these click-based features to identify personal navigation for personalizing the results. Topic-based features extract the topic information of the document, thereby modeling which topics users are interested in. The ODP was a proper tool for representing the topic of documents and was widely used for user modeling~\cite{bennett2010classification, sieg2007web, white2013enhancing}. In order to tackle its incomplete categories and huge labor cost, researches developed the latent topic space to learn the vectors of documents automatically~\cite{Carman2010Towards, harvey2013building, vu2015temporal, vu2017search}. Later studies~\cite{bennett2012modeling, volkovs2015context, white2013enhancing} used the learning to rank method to combine these features.

In recent years, deep learning has been applied to user modeling for personalized search, which is effective in exploring the potential interests of users~\cite{hrnncikm,ZhouDW20,sigir/ZhouDW20,sigir/YaoDW20,cikm/MaDBW20,sigir/ZhouDWXW21,qian2021pchatbot}\dou{give the list of paper here, including zhengyi's cikm paper}\zhou{fixed}. An adaptive ranking model was devised in~\cite{song2014adapting} for building dynamic user profiles. \citet{li2014deep} focused on in-session contextual ranking with semantic features. Recently, various network structures have appeared in personalized search. Recurrent Neural Network was used to model the sequential information of user interests~\cite{hrnncikm}. Generative Adversarial Network was applied to sample high quality training data~\cite{Lu:2019}. Reinforcement Learning was used to model the dynamic change of user search process~\cite{YaoDXW20}. In addition to building user profiles, \citet{sigir/ZhouDW20} and \citet{sigir/YaoDW20} proposed to use the context of history to learn the embedding of the current query. However, although all these existing approaches paid attention to the models for user modeling and personalized ranking, none of them has investigated the use of pre-training tasks for personalized search.\dou{All these existing approaches paid attention to the models for user modeling and personalized ranking, and none of them has investigated the use of pre-training tasks for personalized search.}\zhou{fixed} In this paper, we propose a pre-training\dou{pre-training, can we say pretraining here?}\zhou{maybe we can, but in fact, not the most standard pre-training} framework for personalized search, which is a new paradigm for enhancing data representations.

\begin{figure*}[!t]
	\centering
	\vspace{-0.2cm}
	\setlength{\abovecaptionskip}{0.1cm}
	\includegraphics[width=0.8\linewidth]{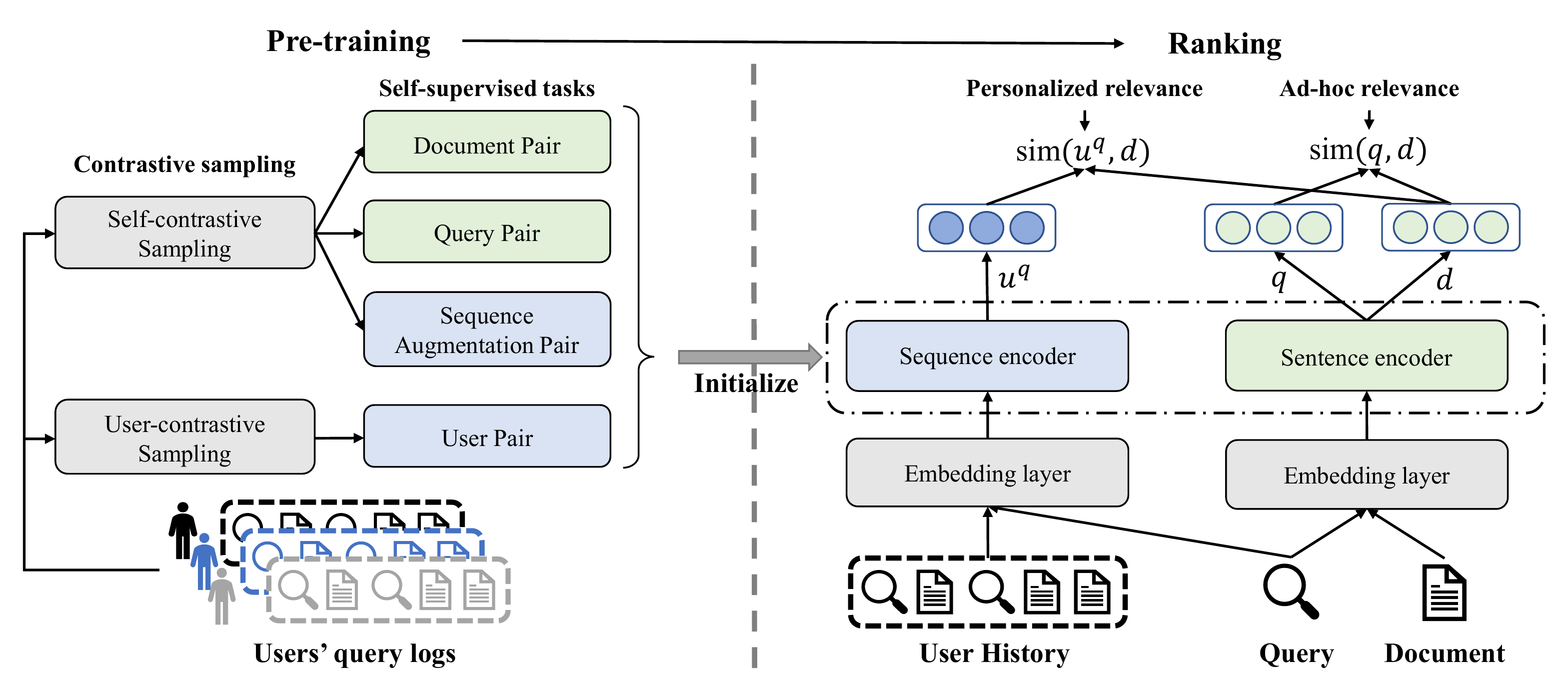}
	\caption{The architecture of our model PSSL, which is a two-stage training model. The first stage is pre-training with four self-supervised tasks that are generated by two angles of contrastive sampling. At the second stage, the sequence encoder and the sentence encoder are initialized with pre-trained parameters. The document ranking score is computed based on them.}
	\label{fig:rank}
\end{figure*}
\subsection{Pre-training for Information Retrieval \dou{pretraining for IR?}\zhou{fixed}}
Self-supervised learning is the mainstream way of pre-training, which uses auxiliary tasks to mine its own supervised information from large-scale unsupervised data. With this constructed supervision information, the network can be pre-trained to learn valuable representations for downstream tasks. Some language models, such as ELMo~\cite{peters2018deep}\dou{please remember to change space to ~ before all cite, ref commands.}\zhou{get it} and BERT~\cite{devlin2018bert}, have made impressive progress in natural language processing. They trained the representations of words or sentences on self-supervised tasks and improved the performance of downstream tasks. In the field of information retrieval, some self-supervised learning frameworks were applied to recommendation. \citet{cikm/ZhouWZZWZWW20} used mutual information maximization to learn the correlations among attributes, items, and sequences. \citet{CL4SR} proposed a contrastive learning framework to learn high-quality user representations. For search tasks, \citet{iclr/ChangYCYK20} designed several paragraph-level pre-training tasks to enhance the representations of queries and documents. \citet{wsdm/MaGZFJC21} presented the model PROP, which constructed fake query-document pairs based on query likelihood model on large-scale corpus. The success of these studies shows that self-supervised learning is able to enhance data representations. In this paper, we attempt to enhance the query, document, and user representations in personalized search with a self-supervised learning framework.

\section{PSSL: our proposed model}\label{sec:method}
Personalized search has become an effective technique to improve user search experience by modeling user interests. As we stated in Section~\ref{sec:intro}\dou{ref, pls check all similar problems in the paper}\zhou{have checked all}, existing personalization models suffer severely from data sparsity and rough data representations. In this paper, we propose a self-supervised learning framework to enhance data representations for personalized search. The architecture of our model PSSL is shown in Figure~\ref{fig:rank}. First, we devise two angles of contrastive sampling to extract self-supervised signals from query logs, and use four tasks for pre-training. And then, the encoders used in the ranking model are initialized with pre-trained parameters. They contribute to high-quality data representations and better personalization.

\subsection{Problem Definition}
Suppose that there is a set of users, denoted as $U$. For each user $u$ in the set, the query log $H_u$ records the user's historical behaviors, including issuing a query and clicking on a document. We represent the user's query log as a sequence, i.e., $H_u=\{q_1,d_{1,1},\cdots,q_{t-1},d_{t-1,1}\}$, where $t$ is the current timestamp, and $d_{i,j}$ refers to the $j_{th}$ clicked document under the query $q_i$. Given the current query $q$, the candidate documents retrieved by the search engine are $\{d_1,d_2,\cdots\}$. Our personalized task is a re-ranking process, which needs to score each of the candidate documents based on the current query and the user's query log. We denote the score of the candidate document $d$ as $\text{score}(d|q,H_u)$, which consists of two parts:
\begin{equation}
\label{eq:score}
    \text{score}(d|q,H_u)=\phi(\text{Pscore}(d|H_u,q),\text{Ascore}(d|q)),
\end{equation}
where $\text{Pscore}(\cdot)$ computes personalized relevance regarding the user history, while $\text{Ascore}(\cdot)$ represents ad-hoc relevance between the current query and the document. The function $\phi(\cdot)$ is a multilayer perceptron with $tanh(\cdot)$ as the activation function.

\subsection{The Architecture of Ranking Model}
As we stated in Section~\ref{sec:intro}\dou{?}\zhou{fixed}, the representation learning tasks in personalized search mainly conclude two dimensions. One is sentence-level encoding, which intends to learn the semantic representations of queries and documents. The other is sequence-level encoding, which focuses on learning user representations from their historical behavior sequences. Based on this consideration, we propose a sentence encoder and a sequence encoder to learn data representations.

The architecture of our ranking model is shown in the right half of Figure~\ref{fig:rank}\dou{?}\zhou{fixed}. First, after the embedding layer, the current query and the document are fed into the sentence encoder to learn the representations. And then, given the user history, the sequence encoder is applied to learn the query-aware user representation based on the current intent. Finally, by matching the document with the current query and the user representation, we can calculate the ranking score of the document and obtain the personalized ranking results. The details of each step are introduced as follows.

\subsubsection{Embedding Layer}
To learn the embedding of each query and document, we initialize a word embedding matrix with $d$-dimension, $M \in \mathbb{R}^{|W|*d}$, where $|W|$ is the vocabulary size. Given a query or a document, we first split it into words and then convert them into word embeddings. The embeddings of the user history are generated by converting each behavior in the sequence into vectors. For convenience, all the characters that appear in the following sections represent the vectors after the word embedding layer. 

\subsubsection{Sentence Encoder}
Previous studies have pointed out that the queries issued by users are usually short and ambiguous~\cite{silverstein1999analysis, cronen2002quantifying}. Moreover, documents retrieved by search engines often contain lots of worthless terms. Therefore, learning accurate semantic representations of queries and documents is integral to search tasks. We design a sentence encoder to tackle this problem. Given the current query $q$ and the candidate document $d$, their sentence-level representations are computed as:
\begin{equation}
        \bar{q}=\text{SenE}(q), \quad \bar{d}=\text{SenE}(d), \nonumber
\end{equation}
where the function $\text{SenE}(\cdot)$ is the sentence encoder. The detailed implementation will be introduced in Section~\ref{sec:implement}.

\begin{figure*}[!t]
	\centering
	\vspace{-0.2cm}
	\setlength{\abovecaptionskip}{0.1cm}
	\includegraphics[width=0.8\linewidth]{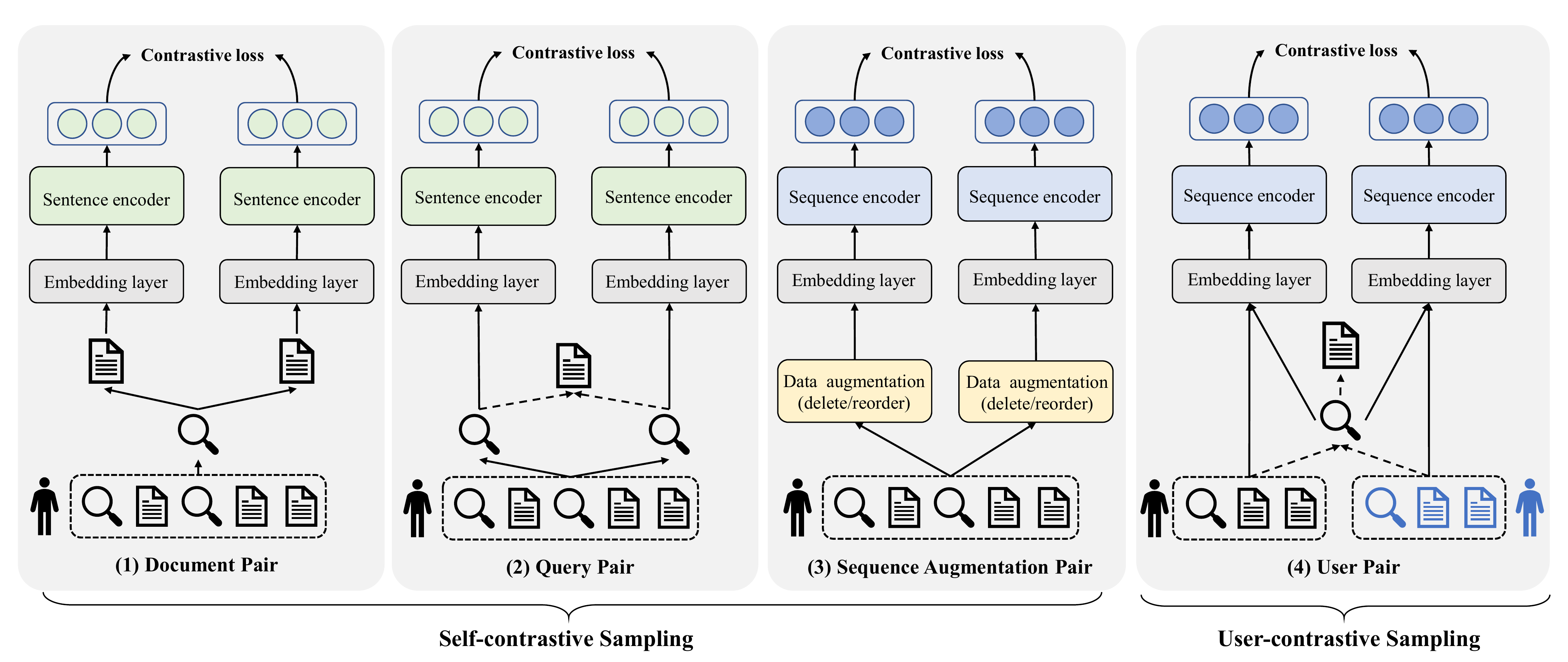}
	\caption{The overview of self-supervised tasks. Under two angles of contrastive sampling, we generate four types of self-supervised signals from query logs, including document pair, query pair, sequence augmentation pair, and user pair. Contrastive loss is used to optimize the parameters of two encoders. The red lines show specific search patterns in query logs.}
	\label{fig:self}
\end{figure*}
\subsubsection{Sequence Encoder}
User representation learning is one of the most crucial tasks in personalized search. A high-quality user representation can accurately reflect the user's preferences, thereby affecting the ranking of the final document list. In general, the user representation is learned from his query log, which can be regarded as a behavior sequence. This promotes us to devise a sequence encoder to model user interests based on historical behaviors. 

Previous studies have shown that a dynamic user profile in response to the current query performs better than the static version~\cite{song2014adapting, hrnncikm}. Inspired by this observation, we feed the user history and the current query together into the sequence encoder to learn the query-aware user representation. Formally, given the concatenated sequence of user's history $H_u$ and current query $q$, the user representation is defined as:
\begin{equation}
        u^q=\text{SeqE}([H_u,q]), \nonumber
\end{equation}
where the function $\text{SeqE}(\cdot)$ is the sequence encoder, which will be also described in Section~\ref{sec:implement}. The user representation $u^q$ will contribute to matching with candidate documents in the following.

\subsubsection{Re-ranking}
As shown in Eq.~(\ref{eq:score})\dou{?}\zhou{fixed}, the final score of candidate documents consists of two parts. For personalized relevance, we compute the similarity between the user representation $u^q$ and the document representation $\bar{d}$:
\begin{equation*}
    \text{Pscore}(d|H_u,q)=\text{Sim}(u^q,\bar{d}), 
\end{equation*}
where the function $\text{Sim}(\cdot)$ is implemented by cosine similarity in this work. For ad-hoc relevance, we first take the similarity between the query representation and the document representation into account. Moreover, following the previous work~\cite{bennett2012modeling, hrnncikm}, we extract additional features to reveal the relevance between the query and the document, including click-based features, topic-based features, and several neural matching features. These features $f_{q,d}$ are fed into MLP to represent the relevance. Finally, the ad-hoc relevance consists of two parts that are combined with another MLP layer\dou{are the two $\phi$s are same or different? you also need to mention $\phi$ is a MLP layer.\zhou{both MLP, different parameters}}:
\begin{equation*}
    \text{Ascore}(d|q)=\phi\left(\text{Sim}\left(\bar{q},\bar{d}\right), \phi(f_{q,d})\right),
\end{equation*}
where $\phi(\cdot)$ represents a multilayer perceptron. Two MLP layers here have different parameters that are learned during the training.

By re-ranking the results based on the final score, we obtain the personalized search results with respect to the user's individual interests. However, due to the shortcomings as we discussed in Section~\ref{sec:intro}\dou{?}\zhou{fixed}, the training of sentence encoder and sequence encoder is limited in data representations. To handle this problem, we apply the self-supervised learning framework to pre-train the two encoders with four auxiliary tasks.

\subsection{Self-supervised Learning with Contrastive Sampling}
Self-supervised learning provides us with sufficient training samples for learning data representations. It can mine the correlations between user behaviors to enhance the generalizability of the model. In this section, we will introduce how to extract self-supervised signals from query logs to pre-train the two encoders.

To design auxiliary tasks that are helpful for personalization, we design the learning objectives considering the characteristic of personalized search. Since that the personalization relies heavily on query logs, we attempt to mine specific search patterns in query logs to construct self-supervised signals. In fact, there are plenty of paired self-supervised information hidden in the query logs. For example, if a user issued ``Online translation'' and ``Google Translate'', and clicked the same URL ``\url{https://translate.google.cn/}'' in the past, we can infer that these two queries reflect the same intent for this user. Therefore, bringing the representations of these two queries closer helps infer the query intent. Such specific search patterns can be transferred to documents and user's behavior sequences for constructing data pairs. Inspired by the contrastive learning framework~\cite{icml/ChenK0H20, cvpr/He0WXG20}, we propose two ways of data sampling to generate paired data, i.e., self-contrastive sampling and user-contrastive sampling. The overview of self-supervised tasks is shown in Figure~\ref{fig:self}. Under two angles of contrastive sampling, four self-supervised tasks are devised to pre-train the encoders.

\subsubsection{Self-contrastive Sampling}
Self-contrastive sampling aims to extract data pairs with similar meanings from a single user's historical behavior sequence. Based on this method, we propose three auxiliary tasks for pre-training.

\textbf{Document Pair}. In the search process, a common scenario is that a user clicks on multiple documents in the returned results under a single query. This indicates that these clicked documents are in line with the user's query intent, and the information contained in these documents has a certain similarity. Based on this consideration, we design the task to close the representations of clicked documents under the same query.

For the user $u$, suppose that he clicks two documents $d_i$ and $d_j$ under the query $q$. After word embedding, we apply the sentence encoder to learn their document representations $\bar{d_i}$ and $\bar{d_j}$:
\begin{equation}
    \bar{d_i}=\text{SenE}(d_i), \quad \bar{d_j}=\text{SenE}(d_j), \quad u \rightarrow q \rightarrow (d_i, d_j), \nonumber
\end{equation}
where the path $u \rightarrow q \rightarrow (d_i, d_j)$ means the user $u$ issued the query $q$ and clicked the documents $d_i$ and $d_j$. The loss function is referred to the contrastive learning loss that maximizes the similarity of positive document pair. Following~\cite{icml/ChenK0H20}, considering a mini-batch of $N$ pairs of documents, we treat $(\bar{d_i}, \bar{d_j})$ as the positive pair and treat other $2(N-1)$ documents within the same mini-batch as negative samples $D^-$. Formally, the loss function of this task $\mathcal{L}_{DP}$ for the document pair $(d_i, d_j)$ can be defined as:
\begin{equation}
    \mathcal{L}_{DP}(d_i,d_j)=-\text{log} \frac{\text{exp}\left(\text{Sim}(\bar{d_i},\bar{d_j})\right)}{\text{exp}\left(\text{Sim}(\bar{d_i},\bar{d_j})\right)+\sum\limits_{d^-\in D^-}\text{exp}\left(\text{Sim}(\bar{d_i},d^-)\right)}, \nonumber
\end{equation}
where the function $\text{Sim}(\cdot)$ is implemented by cosine similarity in this paper. It can also be replaced by inner product.

\textbf{Query Pair}. As we discussed above, users may issue two similar queries to find the same document. This inspires us to mine such a search pattern and to close the representations of paired queries.

Suppose that for the user $u$, he clicked the document $d$ under the query $q_i$ and $q_j$ respectively. These two queries can form a positive pair. The sentence encoder is applied to their embeddings to learn the query representations:
\begin{equation}
    \bar{q_i}=\text{SenE}(q_i), \quad \bar{q_j}=\text{SenE}(q_j), \quad u \rightarrow (q_i,q_j) \rightarrow d. \nonumber
\end{equation}
We also use the contrastive learning framework to train the model. For $N$ pairs of queries in a mini-batch, $(\bar{q_i}, \bar{q_j})$ is regarded as the positive pair, while the others are treated as negative queries $Q^-$. The query pair loss function $\mathcal{L}_{QP}$ for $(q_i, q_j)$ is:
\begin{equation}
    \mathcal{L}_{QP}(q_i,q_j)=-\text{log} \frac{\text{exp}\left(\text{Sim}(\bar{q_i},\bar{q_j})\right)}{\text{exp}\left(\text{Sim}(\bar{q_i},\bar{q_j})\right)+\sum\limits_{q^-\in Q^-}\text{exp}\left(\text{Sim}(\bar{q_i},q^-)\right)}. \nonumber
\end{equation}
The above two pre-training tasks concentrate on learning the parameters of sentence encoder to enhance data representations. To pre-train the sequence encoder, we design two tasks associated with user representations based on their behavior sequences.

\textbf{Sequence Augmentation Pair}. In personalized search, the main task is to model the user interests from his historical behavior sequence. In order to achieve better personalization, we hope that the sequence encoder can highlight the behaviors that best reflect the user's personality. To implement this idea, we apply the sequence augmentation to construct different views of the user's behavior sequence. The representations of augmented sequence pair should be closer than augmented sequences from other users.

Specifically, there are three sequence augmentation strategies for user's behavior sequence. (1) Behavior deleting. This way randomly deletes some behaviors from the sequence to enhance the generalizability of the model. We believe that the remaining behaviors in the sequence can also represent the user to a certain extent. (2) Behavior reorder. This strategy randomly swaps the positions of some behaviors. Although the order of behavior has a certain impact on the modeling of user interests, the user's long-stable preferences will not change and should be highlighted by the sequence encoder. (3) Session deleting. This method randomly deletes some sessions from the sequence. Users sometimes issue a series of queries in one session for a single information need. The remaining behaviors provide a local view of the user representation.

For the user history $H_u$, we apply two random augmentation strategies on the sequence, and get two augmented sequences $H_{u,i}$ and $H_{u,j}$. After the embedding layer, we use the sequence encoder to learn the representations of $H_{u,i}$ and $H_{u,j}$:
\begin{equation}
        s_i=\text{SeqE}(H_{u,i}), \quad s_j=\text{SeqE}(H_{u,j}), \nonumber
\end{equation}
where $s_i$ and $s_j$ are representations of augmented sequences for the user. They respectively represent some of the user preferences, and form the positive pair for training. The augmented sequences from other users' query logs in the same mini-batch are regarded as the negative samples $S^-$. Similarly, the loss function of sequence augmentation pair $\mathcal{L}_{SAP}$ for the user history $H_u$ is:
\begin{equation}
    \mathcal{L}_{SAP}(H_u)=-\text{log} \frac{\text{exp}\left(\text{Sim}(s_i,s_j)\right)}{\text{exp}\left(\text{Sim}(s_i,s_j)\right)+\sum\limits_{s^-\in S^-}\text{exp}\left(\text{Sim}(s_i,s^-)\right)}. \nonumber
\end{equation}

\subsubsection{User-contrastive Sampling}
In addition to self-supervised sampling, we believe that the behavior sequences of different users can also form data pairs. The reason is that users with similar search behaviors tend to have similar interests. Therefore, we attempt to extract self-supervised signals from different users' query logs.

\textbf{User Pair}. For ambiguous queries, the retrieved documents often contain multiple topics. Different user groups tend to click on documents with different topics. For example, for the query ``Apple'', some users prefer ``Apple company'', while others focus on ``Apple fruit''. We believe that users with the same preferences have a certain similarity in user representations. To improve the effectiveness of training, we only choose the queries with ambiguity (click entropy greater than 1.0) to construct user pairs.

For two users $u_i$ and $u_j$, if they both issued the query $q$ and clicked on the same document $d$, these two users can be regarded as the positive pair facing the query $q$. Note that the sequences of user history $H_{u_i}$ and $H_{u_j}$ only contain the behaviors before the query $q$. The sequence encoder is used to learn their query-aware user representations $u^q_i$ and $u^q_j$:
\begin{equation*}
        u^q_i=\text{SeqE}([H_{u_i},q]), \quad u^q_j=\text{SeqE}([H_{u_j},q]), \quad (u_i,u_j) \rightarrow q \rightarrow d.
\end{equation*}
Similar to the previous tasks, the set of negative samples $U^-$ consists of representations of other users in the mini-batch. The user pair loss function $\mathcal{L}_{UP}$ for the user $u_i$ and $u_j$ is:
\begin{equation}
    \mathcal{L}_{UP}(H_{u_i},H_{u_j})=-\text{log} \frac{\text{exp}\left(\text{Sim}(u^q_i,u^q_j)\right)}{\text{exp}\left(\text{Sim}(u^q_i,u^q_j)\right)+\sum\limits_{u^-\in U^-}\text{exp}\left(\text{Sim}(u^q_i,u^-)\right)}. \nonumber
\end{equation}

\subsection{The Implementation Details of Encoders}
\label{sec:implement}
In this section, we will introduce the implementation details of the sentence encoder and the sequence encoder. The basic structure is based on the Transformer encoder~\cite{vaswani2017attention} to model the contextual information following~\cite{sigir/ZhouDW20}.

\textbf{Implementation of Sentence Encoder}. To tackle the query ambiguity and document noise at word level, we attempt to combine contextual information of surrounding words to represent sentences. The transformer encoder is applied to learn the context-aware sentence representations, which is widely used in various fields. Formally, for a query or a document, suppose it consists of $n$ words, i.e., $q=\{w_1,w_2,\cdots,w_n\}$. The sentence encoder is:
\begin{equation}
    \text{SenE}(q)=\text{Transformer}^{\rm sum}_{w}([w_1,w_2,\cdots,w_n]), \nonumber
\end{equation}
where the function $\text{Transformer}^{\rm sum}_{w}$ means the sum of outputs of word-level transformer layer. The final output is the sentence representation which considers the contextual information.

\textbf{Implementation of Sequence Encoder}. Existing studies have pointed out that long-term history and short-term user history play different roles in personalized search~\cite{vu2015temporal, hrnncikm}. Long-term history generally contains user behaviors before the current session, and often reflects the user's long-stable interests. Short-term history refers to the user's past interactions in the current session, which shows the user's recent information needs. Based on this consideration, we use a hierarchical transformer structure to learn user representations following~\cite{sigir/ZhouDW20}.

Given a user history $H_u$, we divide it into long-term and short-term history $H^l_u$ and $H^s_u$, which are fed into two transformers to model user preferences from different views. To learn the user representation based on the current query $q$, we concatenate it with short-term history to model the user's information need of the current session. Moreover, a `[User]' token is added at the end of the sequence to represent the summarized user representation. The sequence encoder consists of a short-term transformer $\text{Transformer}_{s}(\cdot)$ and a long-term transformer $\text{Transformer}_{l}(\cdot)$:
\begin{align}
    \text{SeqE}(H_u,q)&=\text{Transformer}_{l}^{\rm last}\left(\left[H^l_u,u^{q,s}\right]\right), \nonumber \\
    u^{q,s}&=\text{Transformer}_{s}^{\rm last}\left(\left[H^s_u,q,[\text{User}]\right]\right), \nonumber
\end{align}
where the superscript $last$ means taking the last position as the output. Finally, the output represents the user representation which contributes to the personalization of search results.

\subsection{Training and Optimization}
The training process of PSSL includes two stages: pre-training and fine-tuning. At the first stage, we optimize the loss of the four self-supervised tasks ($\mathcal{L}_{DP}$, $\mathcal{L}_{QP}$, $\mathcal{L}_{SAP}$, and $\mathcal{L}_{UP}$) to pre-train the sentence encoder and sequence encoder. At the second stage, two encoders are initialized by the pre-trained parameters, and we use the ranking task to fine-tune them and train the whole network. The ranking loss is computed by cross entropy in a pairwise way:
\begin{equation}
    \mathcal{L}_{Rank}(d_i,d_j)=-\left(\overline{p}_{ij}\text{log}(p_{ij})+\overline{p}_{ji}\text{log}(p_{ji})\right),\nonumber
\end{equation}
where $d_i$ is the positive sample and $d_j$ is the negative sample, $p_{ij}$ and $\overline{p}_{ij}$ represent the predicted probability and the real probability that $d_i$ is more relevant than $d_j$. The $p_{ij}$ is computed by $p(d_i|H_u,q)-p(d_j|H_u,q)$ with sigmoid normalization.

\section{Experimental setup}
\label{sec:experiment}
\subsection{Dataset}
We conduct our experiments on AOL search logs~\cite{pass2006picture} and a commercial dataset. The basic statistics are shown in Table \ref{tab:dataset}. 
\begin{table}[!t]
  \setlength{\belowcaptionskip}{0.1cm}
  \setlength{\abovecaptionskip}{0.1cm}
  \caption{Basic statistics of the datasets.}
  \label{tab:dataset}
  \begin{tabular}{p{0.23\textwidth}p{0.09\textwidth}p{0.09\textwidth}}
  	\toprule
    Dataset & AOL & Commercial \\
  	\midrule
    \# Users & 110,439 & 33,204\\
  	\# Queries & 736,454 & 267,479\\
  	\# Sessions & 279,930 & 97,858\\
  	Average query length & 2.87 & 3.25\\
	Average \#click per query & 1.11 & 1.19\\  
  	\bottomrule
  \end{tabular}
\end{table}

\textbf{AOL dataset:} The AOL search log is a public dataset, which contains users' click-through data from 1st March, 2006 to 31st May, 2006. Following the previous work~\cite{ahmad2018multi}, we remove all non-alphanumeric characters from the queries, and regard the document title as the content to compute the relevance. Since the dataset only records clicked documents without the returned document lists, BM25 algorithm~\cite{robertson2009probabilistic} is used to select candidate documents from the top results. To identify a session, similarity between two consecutive queries is considered with threshold of 0.5. Based on the above operations, each piece of data includes an anonymous user ID, a session ID, a query, the query issued time, a document URL, the ranking position, the document title, and a click tag. Since the personalized search requires historical information to build the basic user profile, we use the first five weeks data as background set. The last eight weeks data is regarded as the experimental set, which is further divided into training set, validation set and test set in a 4:1:1 ratio. To ensure that the data is sufficient for personalization, we remove users whose background set or training set is empty. 

\textbf{Commercial dataset:} This dataset records the query log in January and February, 2013 from a large commercial search engine. There are some differences between this dataset and AOL in data processing. First, the candidate documents are collected directly from the document list returned by the search engine. The original ranking quality is much higher than BM25. Second, we crawl the content corresponding to the URL to represent the document, instead of just the title. This will provide us with more accurate document representations. Third, since this dataset records the click dwell time of each URL, we treat the URL whose click dwell time is greater than 30s as the satisfied document. We keep the same ratio as AOL to collect the background set and the experimental set.

\begin{table*}[!t]
 \center
 \vspace{-0.2cm}
 \setlength{\abovecaptionskip}{0.1cm}
 \setlength{\belowcaptionskip}{0.1cm}
 \caption{The results of all models on two datasets. The percentage is based on the SOTA baseline. `$\dagger$' indicates the model outperforms all baselines significantly with paired t-test at p $<$ 0.05 level. Best results are denoted in bold.}
  \label{tab:overall}
  \begin{tabular}{p{0.1\linewidth}|p{0.033\textwidth}l|p{0.033\textwidth}l|p{0.033\textwidth}l||p{0.033\textwidth}l|p{0.033\textwidth}l|p{0.033\textwidth}l|p{0.033\textwidth}l}
  	\hline
  	\multirow{2}*{Model} & \multicolumn{6}{c||}{AOL dataset}  & \multicolumn{8}{c}{Commercial dataset} \\ \cline{2-15}
  	&\multicolumn{2}{c|}{MAP} & \multicolumn{2}{c|}{MRR} & \multicolumn{2}{c||}{P@1} &\multicolumn{2}{c|}{MAP} & \multicolumn{2}{c|}{MRR} & \multicolumn{2}{c|}{P@1} & \multicolumn{2}{c}{P-improve} \\ \hline
  	\multicolumn{11}{l}{Ad-hoc search baselines} \\ \hline
  	 Ori. & .2504 & -64.9\% & .2596 & -64.2\%  & .1534 & -75.6\% & .7399 & -10.2\% & .7506 & -10.0\% & .6162 & -15.6\% & - & -\\
  	KNRM & .4291 & -39.8\% & .4391 & -39.5\%  & .2704 & -56.9\% & .4916 & -40.3\% & .5001 & -40.1\% & .2849 & -61.0\% & .0655 & -75.3\% \\
  	Conv-KNRM & .4738 & -33.5\% & .4849 & -33.2\% & .3266 & -48.6\% & .5872 & -28.7\% & .5977 & -28.4\% & .4188 & -42.7\% & .1422 & -46.5\%\\
  	BERT & .5033 & -29.4\% & .5135 & -29.3\% & .3552 & -43.4\% & .6232 & -24.4\% & .6326 & -24.2\% & .4475 & -38.7\% & .1778 & -33.1\%\\ \hline
  	\multicolumn{11}{l}{Personalized search baselines} \\ \hline
	HRNN & .5423 & -23.9\% & .5545 & -23.6\% & .4854 & -22.7\% & .8065 & -2.1\% & .8191 & -1.8\% & .7127 & -2.4\% & .2404 & -9.5\%\\
	PSGAN & .5480 & -23.1\% & .5601 & -22.8\% & .4892 & -22.1\% & .8135 & -1.3\% & .8234 & -1.3\% & .7174 & -1.8\% & .2489 & -6.3\%\\
	RPMN & .5926 & -16.9\% & .6049 & -16.7\% & .5322 & -15.2\% & .8238 & - & .8342 & - & .7305 & - & .2656 & - \\
    HTPS & .7091 & -0.5\% & .7251 & -0.1\% & .6268 & -0.1\% & .8224 & -0.2\% & .8324 & -0.2\% & .7286 & -0.3\% & .2552 & -3.9\%\\
    PEPS & .7127 & - & .7258 & - & .6279 & - & .8221 & -0.2\% & .8321 & -0.3\% & .7251 & -0.7\% & .2545 & -4.2\% \\ \hline
	\multicolumn{11}{l}{Our method} \\ \hline
	PSSL & $\bm{.7359}^\dagger$ & +3.3\% & $\bm{.7484}^\dagger$ & +3.1\% & $\bm{.6431}^\dagger$ & +2.4\% & $\bm{.8301}^\dagger$ & +0.8\% & $\bm{.8398}^\dagger$ & +0.7\% & $\bm{.7338}^\dagger$ & +0.5\% & $\bm{.2688}^\dagger$ & +1.2\%\\
    \hline
  \end{tabular}
\end{table*}
\subsection{Baselines}
To compare the performance of our model and other neural ranking models, we select several typical ad-hoc search models and personalized search models as baselines. They are:

\textbf{KNRM}~\cite{xiong2017end}. It is a neural ranking model based on kernel-pooling for ad-hoc search. Multi-level soft matching features are extracted from the word similarity matrix for ranking.

\textbf{Conv-KNRM}~\cite{dai2018convolutional}. It adds a convolutional layer on the KNRM to model n-gram soft matches. Contextual information of surrounding words are considered to improve matching accuracy. 

\textbf{BERT}~\cite{qiao2019understanding}. This model applies the pre-trained BERT model to query-document matching task. The concatenated query-document sequence is fed into the pre-trained BERT model. The last layer's representation of `[CLS]' token is regarded as the matching features. The BERT model are fine-tuned during the training.

\textbf{HRNN}~\cite{hrnncikm}. For personalized search, this work models the sequential information of query logs and learns dynamic user profiles based on the current query. Hierarchical recurrent neural networks with query-aware attention is used to implement this idea.

\textbf{PSGAN}~\cite{Lu:2019}. This study concentrates on data augmentation based on generative adversarial network for personalized search. It aims to extract valid training data from limited and noisy click data. Considering the cost of training, we take the discriminator in the document selection based model as the baseline.

\textbf{RPMN}~\cite{ZhouDW20}. This is a memory network-based personalized search model, which attempts to identify potential re-finding behaviors in personalized search. It devises three external memories to cover two types of re-finding behavior.

\textbf{PEPS}~\cite{sigir/YaoDW20}. This model trains personal word embeddings for each user based on his historical data, and abandons the construction of user profiles. Personal and global word embeddings are both considered for better data representations.

\textbf{HTPS}~\cite{sigir/ZhouDW20}. This is a personalized search framework based on hierarchical transformer. Transformer encoder is first used to encode history as contextual information to disambiguate the query.

\subsection{Implementation Details}
For our proposed model PSSL\footnote{The code of the model is available on https://github.com/smallporridge/PSSL.}, the word embedding matrix are initialized by the word2vec~\cite{mikolov2013exploiting} model following~\cite{hrnncikm, Lu:2019}. It will be fixed in the pre-training phase, and be fine-tuned during the training of ranking task. We conducted multiple experiments to select the parameters of the model. Finally, the dimension of the word embedding is 100. The hidden size of transformer is 512. The number of attention heads in transformer is 6. The number of transformer layers is 6. The number of MLP hidden units is 128. The learning rate of the pre-training task and the ranking task are set to $1e^{-3}$ and $3e^{-4}$ respectively. At the pre-training stage, for sequence augmentation strategies, we change 50\% of user behaviors in the sequence. For four tasks (DP, QP, SAP, and UP), we sample about 163k, 293k, 728k, 128k contrastive pairs on the AOL dataset and 52k, 103k, 264k, 21k contrastive pairs on the commercial dataset. The weights for the four losses ($\mathcal{L}_{DP}$, $\mathcal{L}_{QP}$, $\mathcal{L}_{SAP}$, and $\mathcal{L}_{UP}$) are set as 0.5, 0.5, 1.0, 0.2.

\subsection{Evaluation Metrics}
Since the AOL dataset does not contain user click dwell time, we simply label 
the clicked documents as relevant, while the satisfied documents in the commercial dataset are regarded as relevant. To evaluate the model performance, we employ mean average precise(MAP), mean reciprocal rank (MRR), and precision@1 (P@1) to measure the ranking quality. However, the above metrics are somewhat problematic due to the position bias~\cite{Joachims2005Accurately} in re-ranking tasks. To measure the ranking results in a more objective manner, we apply another metric called P-improve to evaluate reliable improvements on the inverse document pair following previous works~\cite{hrnncikm, Lu:2019}. Since the candidate documents of AOL dataset are not presented to users, we only use this metric on the commercial dataset which suffers from the position bias.

\begin{table*}[!t]
 \center
 \vspace{-0.2cm}
 \setlength{\abovecaptionskip}{0.1cm}
 \setlength{\belowcaptionskip}{0.1cm}
 \caption{Performance of ablation studies on self-supervised tasks of the PSSL model.}
  \label{tab:ablation}
  \begin{tabular}{p{0.12\linewidth}|p{0.033\textwidth}l|p{0.033\textwidth}l|p{0.033\textwidth}l||p{0.033\textwidth}l|p{0.033\textwidth}l|p{0.033\textwidth}l|p{0.033\textwidth}l}
  	\hline
  	\multirow{2}*{Model} & \multicolumn{6}{c||}{AOL dataset} & \multicolumn{8}{c}{Commercial dataset} \\ \cline{2-15}
  	&\multicolumn{2}{c|}{MAP} & \multicolumn{2}{c|}{MRR} & \multicolumn{2}{c||}{P@1} &\multicolumn{2}{c|}{MAP} & \multicolumn{2}{c|}{MRR} & \multicolumn{2}{c|}{P@1} & \multicolumn{2}{c}{P-improve} \\ \hline
  	\multicolumn{15}{l}{Tasks of sentence encoder} \\ \hline
	\;\;w/o. DP & .7296 & -0.9\% & .7434 & -0.7\% & .6389 & -0.7\% & .8296 & -0.1\% & .8390 & -0.1\% & .7331 & -0.1\% & .2678 & -0.4\%\\
	\;\;w/o. QP & .7252 & -1.5\% & .7398 & -1.1\% & .6368 & -1.0\% & .8288 & -0.2\% & .8381 & -0.2\% & .7325 & -0.2\% & .2662 & -1.0\%\\
	\;\;w/o. DP+QP & .7200 & -2.2\% & .7357 & -1.7\% & .6341 & -1.4\% & .8287 & -0.2\% & .8381 & -0.2\% & .7324 & -0.2\% & .2658 & -1.1\%\\ \hline
	\multicolumn{15}{l}{Tasks of sequence encoder} \\ \hline
	\;\;w/o. SAP & .7212 & -2.0\% & .7362 & -1.6\% & .6343 & -1.4\% & .8242 & -0.7\% & .8348 & -0.6\% & .7288 & -0.7\% & .2584 & -3.9\%\\
    \;\;w/o. UP & .7288 & -1.0\% & .7430 & -0.7\% & .6382 & -0.7\% & .8270 & -0.4\% & .8366 & -0.4\% & .7306 & -0.4\% & .2632 & -2.1\%\\
    \;\;w/o. SAP+UP & .7170 & -2.6\% & .7330 & -2.1\% & .6316 & -1.8\% & .8222 & -1.0\% & .8321 & -1.0\% & .7272 & -0.9\% & .2548 & -5.2\%\\ \hline
	PSSL & .7359 & - & .7484 & - & .6431 & - & .8301 & - & .8398 & - & .7338 & - & .2688 & -\\
    \hline
  \end{tabular}
\end{table*}
\section{Results and Analysis}\label{sec:result}
\subsection{Overall Performance Comparison}
The results of different models on the two datasets are shown in Table~\ref{tab:overall}. It can be observed that:

(1) Our method vs. baselines. Our proposed model PSSL outperforms all baseline models on both datasets. Compared with the best baseline model, PSSL shows significant improvements in all evaluation metrics with paired t-test at p $<$ 0.05 level. Specifically, our model improves the ranking quality by 3.3\% on MAP on the AOL dataset compared with PEPS, while outperforms RPMN by 0.8\% on the commercial dataset. This indicates that the self-supervised learning framework helps the model learn better data representations and further improves the personalization. The improvement on the two datasets also verifies the generalizability of our model. 

(2) Personalized search vs. ad-hoc search. The ad-hoc search concentrates on improving matching accuracy between the query and the document, while the personalized search focuses on how to model user interests based on historical behaviors. All neural personalized search models outperform ad-hoc baselines significantly, while the improvement on the metric P@1 is more obvious. This reflects the personalization models perform well on modeling user's re-finding behavior. Our model PSSL combines the advantages of these two types of search models, using a sentence encoder for query-document matching and a sequence encoder for user representation learning. With a self-supervised learning framework, this is proven to be effective in improving the ranking quality.

(3) AOL dataset vs. commercial dataset. The commercial dataset has relatively higher original ranking quality than AOL, which leads to the results that ad-hoc search baselines perform worse than the original ranking on the commercial dataset. The model HTPS and PEPS, which incorporate interactive matching features between the query and the document, make a significant improvement on the AOL dataset, but they underperform the RPMN on the commercial dataset. This shows that the AOL dataset tests the model on modeling user interests and text matching at the same time, while the commercial dataset focuses more on testing the personalization capabilities of the model. Our model performs well on both datasets, which further proves the robustness of PSSL model.

In summary, the results indicate that \textbf{self-supervised learning with contrastive sampling for personalized search is conducive to refine data representations and promote search results personalization.} To test the model in more detail, we conduct several supplementary experiments: ablation studies, effect of self-supervised learning, and performance on different query sets.

\subsection{Ablation Studies}
To verify the necessity of each of our self-supervised tasks, we conduct ablation experiments on two datasets for the whole model PSSL. Specifically, we explore the role of each self-supervised task on the two encoders respectively, including the tasks of document pair (DP), query pair (QP), sequence augmentation pair (SAP), and user pair (UP). We also remove the pre-training of sentence encoder (DP+QP) or sequence encoder (SAP+UP) to observe the results.

As shown in Table~\ref{tab:ablation}, the removal of each self-supervised task will damage the results on all evaluation metrics. Concretely, deleting the task of SAP causes the most obvious impact on performance on both datasets. This indicates that our sequence augmentation strategies help the sequence encoder model the user representations more accurately. Meanwhile, the task of UP also makes some contributions to the results, which shows that closing the distance between similar users is useful for user modeling. Additionally, we find that removing the pre-training of sentence encoder causes a severe drop on the AOL dataset, while it has little effect on the commercial dataset. A possible reason is that the pre-training of sentence encoder is more helpful for computing ad-hoc relevance. For the AOL dataset, the ad-hoc relevance is more useful due to the poor original ranking quality. But for the commercial dataset, its high-quality original ranking results have provided effective ad-hoc relevance. It can be seen that the task of QP is more important than DP for personalizing the results. This indicates that the queries can provide more additional personalized information for the model.

\begin{figure}[!t]
\vspace{-0.3cm}
    \begin{subfigure}{.49\linewidth}
    \centering
    \includegraphics[width=\linewidth]{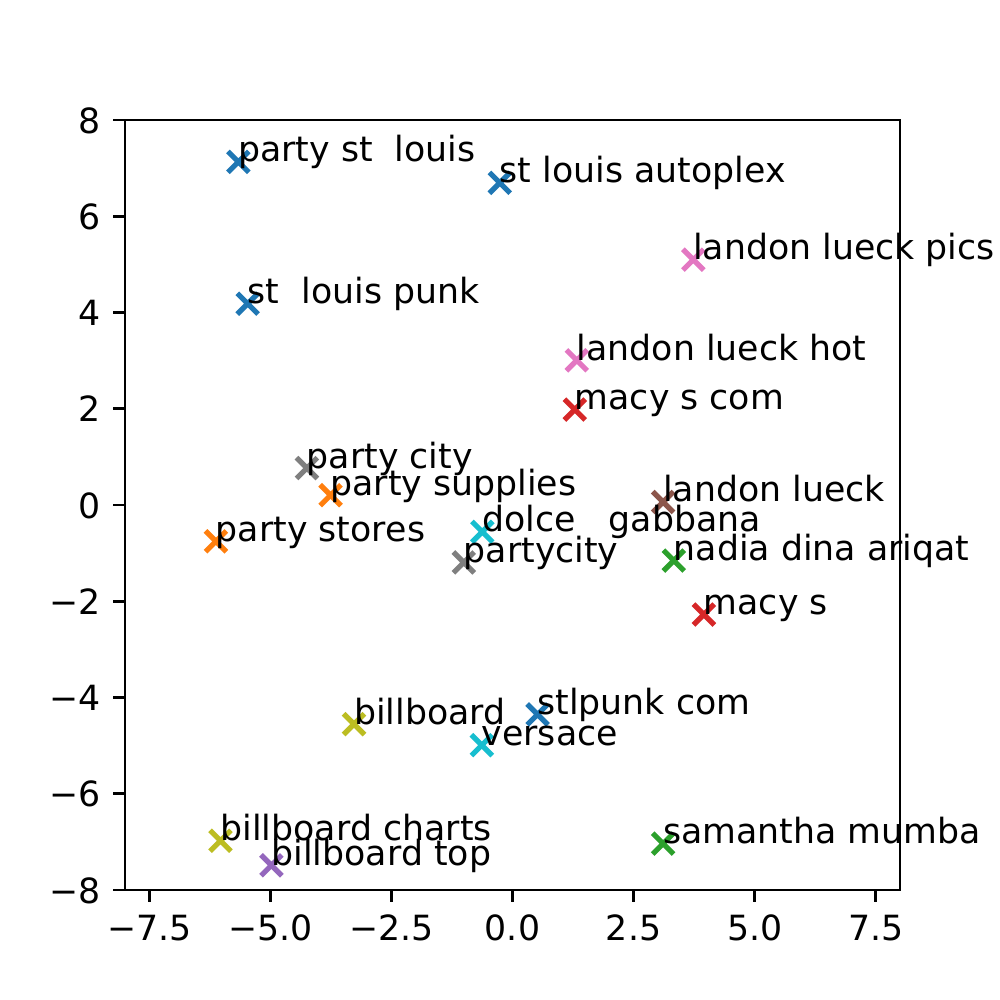}
    \caption{Initialized}
    \end{subfigure}
    \begin{subfigure}{.49\linewidth}
    \centering
    \includegraphics[width=\linewidth]{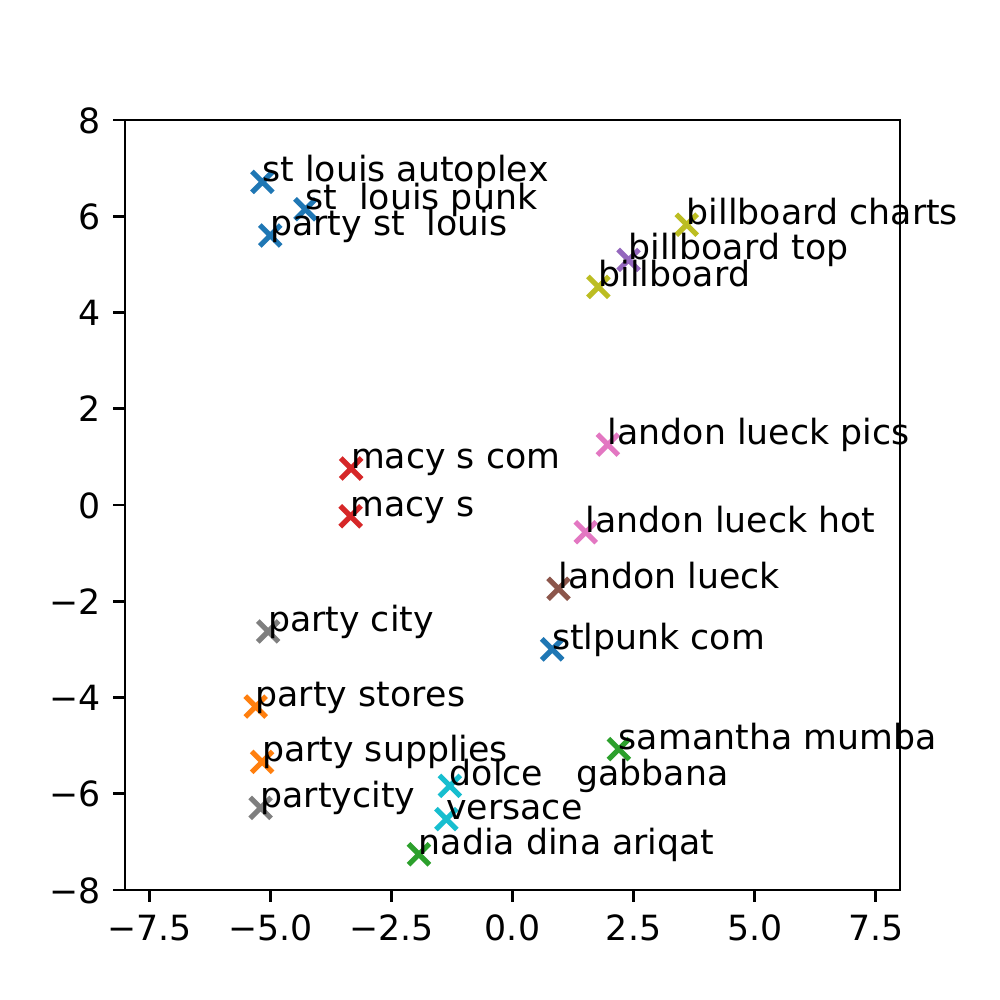}
    \caption{Self-supervised}
    \end{subfigure}
    \caption{The query distribution of the user \#104.} 
    \label{fig:sene}
\end{figure}
\subsection{Effect of Self-supervised Learning}
In order to explore the impact of self-supervised tasks on data representations in more detail, we visualize the quality of the sentence encoder and sequence encoder respectively.

\textbf{Quality of Sentence Encoder.} The self-supervised tasks for sentence encoder mainly enhance the representations of queries and documents, so as to close the distance between similar data. In order to verify the enhancement of data representations by the self-supervised tasks, we compare the distribution of initialized and self-supervised query vectors. Specifically, we randomly select a user from the AOL dataset and map his high-dimensional query vectors to a two-dimensional space through PCA. Queries containing the same clicked document are set to the same color.

The results are shown in Figure~\ref{fig:sene}. We find the initialized query distribution is more dispersive, and self-supervised learning makes the distance of queries with the same color closer. For instance, the queries with red color ``macy s'' and ``macy s com'' have a certain distance in the initialized distribution, but are obviously closer after the self-supervised learning. This indicates that although these two queries have some differences in words, they tend to reflect the same query intent for the user in search scenario. Additionally, the self-supervised learning intends to map the queries into multiple groups. The distance within the group is close, while the boundary between the groups is relatively clear. This shows our model can not only close the distance between similar queries, but also widen the distance between different groups.

\textbf{Quality of Sequence Encoder.} The sequence encoder is designed for learning accurate user representations, which is a critical indicator to distinguish users. In order to observe the impact of self-supervised tasks on the sequence encoder, we compare the difference between initialized and self-supervised user representations. Specifically, we randomly select 1000 users from the AOL dataset and calculate the similarity between the representations of every two users. We divide the bar at 0.025 intervals, and count the number of user pairs that fall in each range.
\begin{figure}[!t]
	\centering
	\vspace{-0.05cm}
	\setlength{\abovecaptionskip}{0.1cm}
	\includegraphics[width=0.98\linewidth]{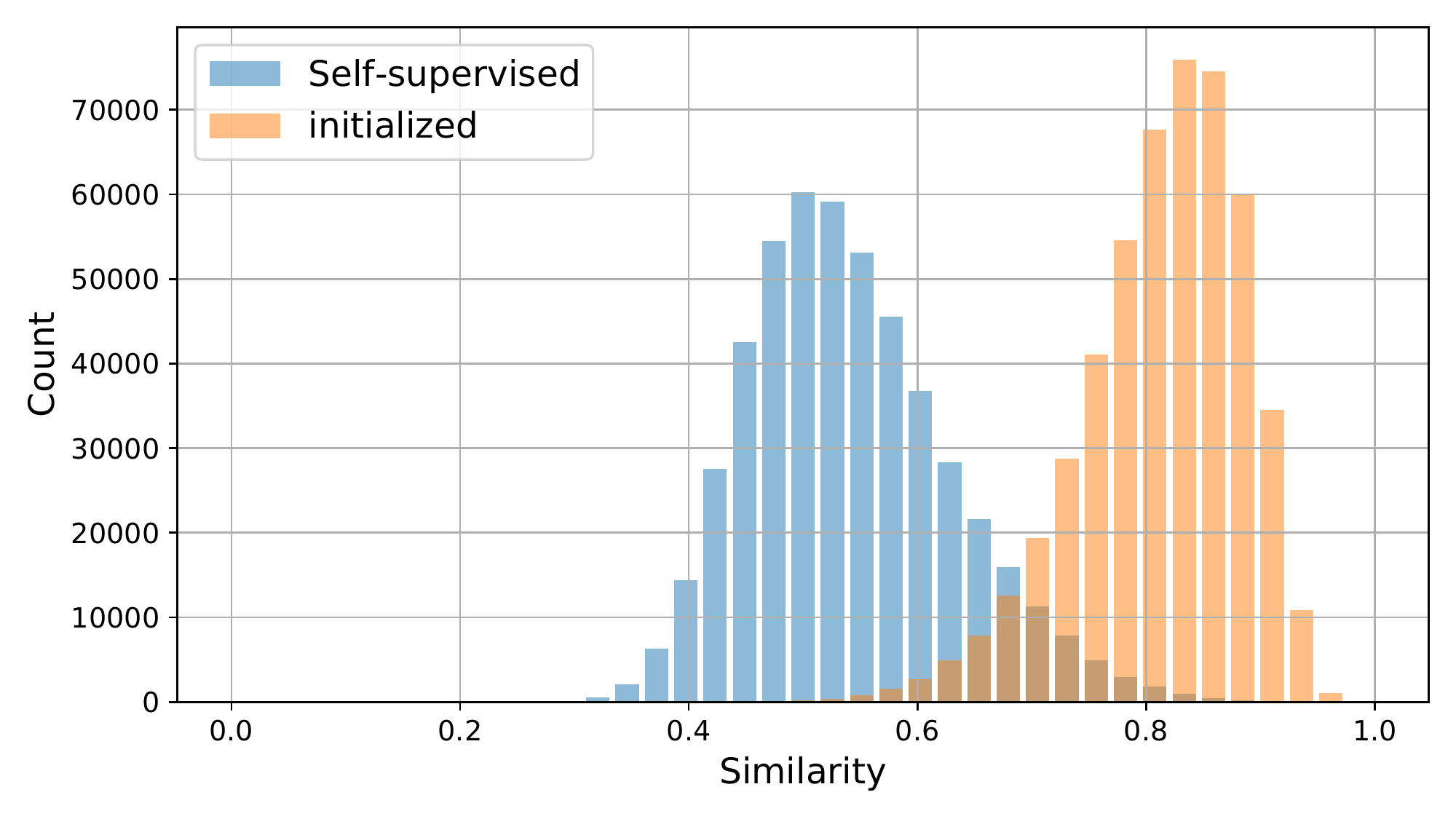}
	\caption{Distribution of similarity between users.}
	\label{fig:seqe}
\end{figure}

From Figure~\ref{fig:seqe}, it can be seen that the similarity between user representations generally obeys normal distribution. The initialized user representations show higher consistency, which means that the model cannot effectively identify the differences between users. With the self-supervised tasks, the average similarity between users becomes smaller. This indicates that the differences between users are magnified. Another interesting finding is that for the self-supervised user representations, the distribution is not completely symmetrical. Ranges with greater similarity contain more user pairs. This may benefit from the self-supervised task of UP, which aims to close the distance between similar users.

\subsection{Performance on Different Query Sets}
In the search process, according to the different search purposes, the user's queries can be classified as the navigational query and the informational query. For navigational queries, different users tend to have the same query intent. The informational queries are usually ambiguous and have multiple meanings. To measure the query ambiguity, we compute the click entropy and set 1.0 as the threshold to divide the queries into two subsets. We choose the best baseline model for comparison. Additionally, we remove the pre-training of sentence encoder (w/o. SenE) or sequence encoder (w/o. SeqE) for detailed analysis. We use the improvement of MAP over original ranking to show the performance.

\begin{figure}[!t]
    \vspace{-0.1cm}
    \begin{subfigure}{.49\linewidth}
    \centering
    \includegraphics[width=\linewidth]{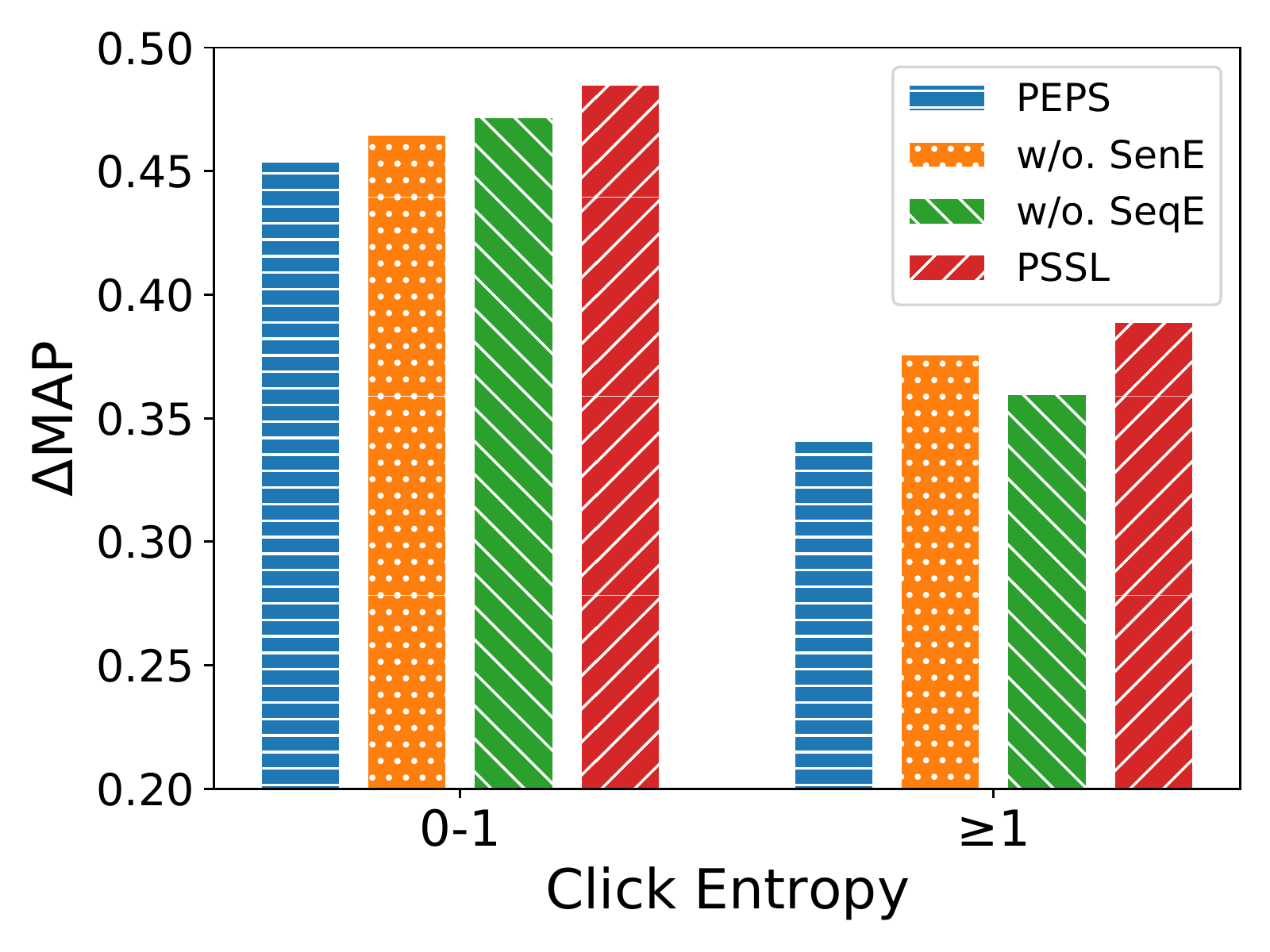}
    \caption{AOL dataset}
    \end{subfigure}
    \begin{subfigure}{.49\linewidth}
    \centering
    \includegraphics[width=\linewidth]{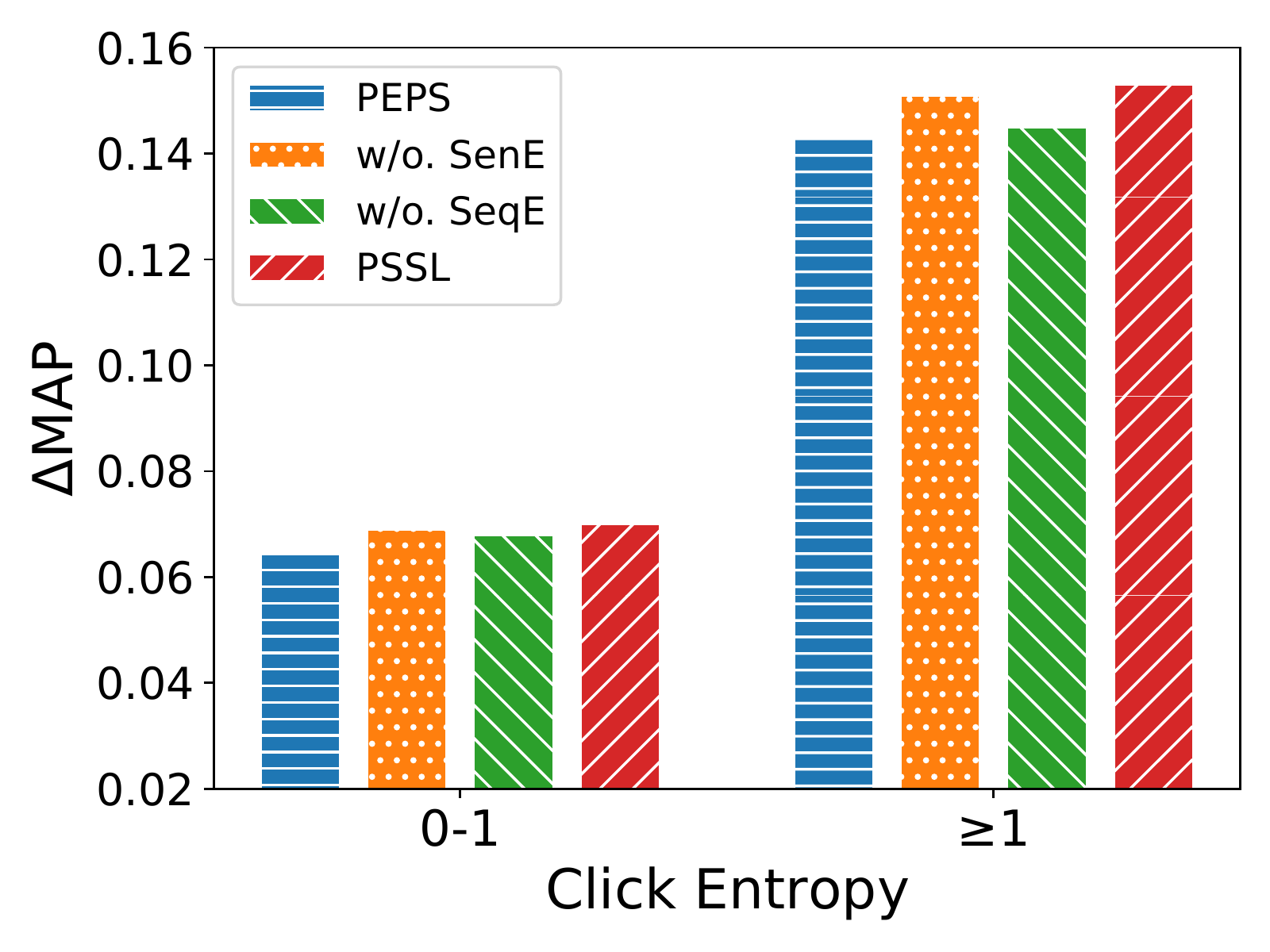}
    \caption{Commercial dataset}
    \end{subfigure}
    \caption{The results on queries with different click entropies with the threshold at 1.0.} 
    \label{fig:entropy}
\end{figure}

As shown in Figure~\ref{fig:entropy}, all models perform better on navigational queries for the AOL dataset, but this is inconsistent with the results on the commercial dataset. A possible reason is that the commercial dataset has a high-quality original ranking, which has little room for improvement on navigational queries. Our proposed model PSSL outperforms PEPS on both query sets, especially on the queries with larger click entropy. This indicates that our model is able to learn high-quality user representations when facing ambiguous queries. Specifically, removing the pre-training of sequence encoder causes severe decline on informational queries. This shows that the pre-trained sequence encoder contributes to modeling user interests more. For sentence encoder, the self-supervised tasks show effectiveness on the AOL dataset, but the contribution on the commercial dataset is limited. This is in line with the characteristics of the two datasets. 

\section{Conclusion}\label{sec:conclusion}
In this paper, we proposed a self-supervised learning framework for personalized search to enhance data representations. First, we presented a ranking model which consists of a sentence encoder and a sequence encoder. Next, we designed two angles of contrastive sampling methods to generate paired self-supervised data from users' query logs. Four auxiliary tasks were devised to pre-train the two encoders for personalized search. Endowed with the benefit of pre-trained parameters, we could get better data representations to improve the personalized results and the generalizability of the model. Experimental results confirmed the effectiveness and robustness of our proposed two-stage training framework.

\begin{acks}
Zhicheng Dou is the corresponding author. This work was supported by Shandong Provincial Natural Science Foundation (No. ZR2019ZD06), National Natural Science Foundation of China (No. 61872370 and No. 61832017),  Beijing Outstanding Young Scientist Program (No. BJJWZYJH012019100020098), the Outstanding Innovative Talents Cultivation Funded Programs 2020 of Renmin Univertity of China, and Intelligent Social Governance Platform, Major Innovation \& Planning Interdisciplinary Platform for the ``Double-First Class'' Initiative, Renmin University of China. We also wish to acknowledge the support provided and contribution made by Public Policy and Decision-making Research Lab of Renmin University of China.
\end{acks}

\balance
\bibliographystyle{ACM-Reference-Format}
\bibliography{references}
\end{document}